\documentclass[conference]{IEEEtran}
\IEEEoverridecommandlockouts
\usepackage{cite}
\usepackage{amsmath,amssymb,amsfonts}
\usepackage{algorithmic}
\usepackage{graphicx}
\usepackage{textcomp}
\usepackage{xcolor}
\usepackage{siunitx}
\usepackage{booktabs}
\usepackage{acronym}
\usepackage{hyperref}
\usepackage{nicefrac}
\usepackage[ruled,vlined,titlenumbered,linesnumbered]{algorithm2e}
\usepackage{balance}
\usepackage{pifont}
\usepackage{subcaption}
\usepackage{epstopdf}
\newcommand{\xmark}{\ding{55}}%

\def\BibTeX{{\rm B\kern-.05em{\sc i\kern-.025em b}\kern-.08em
    T\kern-.1667em\lower.7ex\hbox{E}\kern-.125emX}}
\newcommand{\refeq}[1]{(\ref{#1})}
\newcommand{\reffig}[1]{Fig.~\ref{#1}}
\newcommand{\refsection}[1]{Section~\ref{#1}}
\newcommand{\reftable}[1]{Table~\ref{#1}}

\newcommand{\refalgo}[1]{Algorithm~\ref{#1}}

\acrodef{BLE}{Bluetooth Low Energy}
\acrodef{UUID}{Universally Unique Identifier}
\acrodef{DoD}{direction of departure}
\acrodef{DoA}{direction of arrival}
\begin{document}

\title{Identifying the BLE Advertising Channel for Reliable Distance Estimation on Smartphones}

\author{\IEEEauthorblockN{Christian Gentner\textsuperscript{*}\thanks{\textsuperscript{*}All authors contributed equally to this paper. This work has been submitted to the IEEE for possible publication. Copyright may be transferred without notice, after which this version may no longer be accessible.}}
\IEEEauthorblockA{
\textit{German Aerospace Center (DLR)}\\
Wessling, Germany\\
christian.gentner@dlr.de}
\and
\IEEEauthorblockN{Daniel G\"unther\textsuperscript{*}}
\IEEEauthorblockA{ \textit{Technical University of Munich (TUM)}\\
Munich, Germany \\
d.guenther@tum.de}
\and
\IEEEauthorblockN{Philipp H. Kindt\textsuperscript{*}}
\IEEEauthorblockA{
\textit{Technical University of Munich (TUM)}\\
Munich, Germany \\
philipp.kindt@tum.de}
}


\maketitle
 \IEEEpubidadjcol
\begin{abstract}
As a response to the global COVID-19 surge in 2020, many countries have implemented lockdown or stay-at-home policies. If, however, the contact persons of every infected patient could be identified, the number of virus transmissions could be reduced, while the more incisive measures could be softened. For this purpose, contact tracing using smartphones is being considered as a promising technique. Here, smartphones emit and scan for Bluetooth Low Energy (BLE) signals for detecting devices in range. When a device is detected, its distance is estimated by evaluating its received signal strength. The main insight that is exploited for distance estimation is that the attenuation of a signal increases with the distance along which it has traveled. However, besides distance, there are multiple additional factors that impact the attenuation and hence disturb distance estimation. Among them, frequency-selective hardware and signal propagation belong to the most significant ones. For example, a BLE device transmits beacons on three different frequencies (channels), while the transmit power and the receiver sensitivity depend on the frequency. As a result, the received signal strength varies for each channel, even when the distance remains constant. However, the information on which wireless channel a beacon has been received is not made available to a smartphone. Hence, this error cannot be compensated, e.g., by calibration. In this paper, we for the first time provide a solution to detect the wireless channel on which a packet has been received on a smartphone. We experimentally evaluate our proposed technique on multiple different smartphone models. Our results help to make contact tracing more robust by improving the accuracy of distance estimation.
\end{abstract}
\maketitle

\section{Introduction}
The global surge of the novel coronavirus SARS-CoV-2 has resulted in many countries implementing lockdown and stay-at-home policies. This has lead to a burdensome situation for the population and causes severe economic problems. 
However, whenever a person is tested positive, if all of their contacts could be identified, the further spread of the virus could potentially be stopped or slowed down, while other, more incisive measures could be softened. Thereby, it is of crucial importance that contact persons are identified and warned at the earliest possible time.

For this purpose, smartphone-based contact tracing approaches are being considered as an important tool. Here, every device continually transmits \ac{BLE} beacons and listens for incoming transmissions. As soon as a beacon from another device is received, the distance between both of them is estimated. When the contact duration (which is estimated based on the period of time during which beacons from a particular device are received) is sufficient and/or the estimated distance is small enough, a contact is considered as relevant. Hence, when a person is tested positive for SARS-CoV-2, all relevant contacts can be identified in retrospect by evaluating the smartphone data~\cite{gnther2020tracing}. Besides contact tracing, distance- or proximity estimation plays a key role in a host of applications, such as indoor navigation or object tracking.\\

\par\noindent\textbf{Distance Estimation: }Distance estimation works as follows. One device transmits a beacon with a certain transmit power $P_t$. This value is piggy-backed onto the beacon. The wireless signal then undergoes a certain \textit{path loss}, which depends on the distance along which the signal travels. 
The opposite device will receive the beacon with a certain power $P_r$.
In free space, a wireless signal traveling along a distance $d$ between a sender and a receiver will be received with a power of
\begin{equation}\label{eq:pathloss}
P_r = P_t \cdot G_t G_r \cdot \left(\frac{\lambda}{4 \pi d}\right)^2\,,
\end{equation}
see~\cite{Fri46}. In Equation \eqref{eq:pathloss}$, \lambda$ is the wavelength of the signal and $G_t$ and $G_r$ are the gains of transmitter and receiver, respectively. $G_t$ and $G_r$ can be obtained by calibrating each smartphone model individually. Since $P_r$ and $P_t$ are known by the receiver, the distance $d$ can be estimated.

This estimation is aggravated by multiple effects, of which the most important ones are the following.
\begin{itemize}
    \item The antennas of both devices might be directional and hence, the orientation between both devices impacts $P_t$ and $P_r$. 
    \item Human tissue attenuates the signal by a considerably higher degree than free space. For example, the attenuation between the chest and the back of the human body has been reported as $\SI{19.2}{dB}$~\cite{alomainy:07}.
    As a result, the estimated distance is strongly disturbed when human tissue is within the direct signal path, and hence also by how two human bodies are oriented relatively to each other, and where the phones are worn on the body.
    \item Multipath propagation, e.g., reflections on metal surfaces, can interfere with the direct signal.
    \item The sensitivity of $G_t$ and $G_r$ depend on the wireless channel and therefore the frequency on which a beacon is received.
\end{itemize}
On smartphones, the received power of a BLE signal is available in the form of a received signal strength indicator (RSSI), which is provided by the Bluetooth radio.
Distance estimation using the RSSI on smartphones has been studied thoroughly throughout the last years, e.g., in \cite{chowdhury:15, liu:11}. Currently, due to the high relevance for contact tracing, analyzing and increasing the accuracy of distance estimation is receiving considerable attention by the scientific community~\cite{sattler:20, leith:20}.\\

\par\noindent\textbf{Channel-Dependent RSSI: }This paper addresses the problem of the frequency dependence of the RSSI, which reduces the accuracy of distance estimation. In the BLE protocol, which is used for contact tracing on smartphones, advertising packets are sent on 3 different channels, which are spread over almost the entire frequency band used. 
These 3 channels use center frequencies of $\SI{2.402}{GHz}, \SI{2.426}{GHz}$ and $\SI{2.480}{GHz}$~\cite{bleSpec52}. 
Because of the following three effects, the RSSI depends on the channel on which the packet was received.
\begin{enumerate}
\item Almost every device has frequency-dependent values of $G_r$ and $G_t$. In other words, a packet sent on a certain channel (and hence frequency) will have a larger power than when being sent on a different channel, and the receiver will similarly sense different RSSI values for the same actual signal power on different channels. This effect can lead to a RSSI estimation error of multiple dB, which can result in a distance estimation error of several meters. 
If it is known on which channel each packet is received, then $G_r$ and $G_t$ can be measured separately for each channel, and hence the error can be cancelled easily.
\item The path loss of a signal depends on the channel/frequency on which the signal is transmitted, see Equation \eqref{eq:pathloss}. This effect can also be cancelled easily when the channel on which a packet was received is known. 
\item Packets sent on different channels propagate differently in the environment. BLE signals are reflected, scattered and diffracted by objects in the surrounding. 
Hence, the signal reaching the receiving antenna consists of multiple replicas of the transmitted signal, which are called multipath components. 
These replica of the transmitted signal interfere with those transmitted along the direct path, i.e., the line-of-sight path.
When interfering constructively, the RSSI increases, whereas it is reduced in the case of destructive interference.
Thus, the RSSI accuracy might be drastically reduced due to the distorted received signal by multipath propagation. 
\end{enumerate}
\begin{figure}[tb]
{
  \centerline{\includegraphics[width=0.99\linewidth]{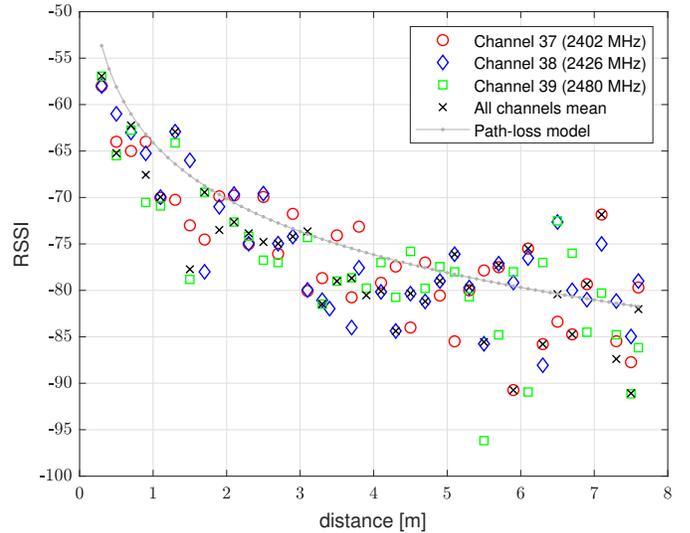}}
    \caption{Measured RSSI at a Google Pixel 3 smartphone for different distances. As a transmitter a Google Pixel 2 was used. }
    \label{fig:RecordedDataDiffDistances}
 }
 \end{figure}

These effects are shown in \reffig{fig:RecordedDataDiffDistances}, which depicts multiple RSSI measurements for the 3 different channels used in BLE. Here, we used a Google Pixel 2 smartphone as a transmitter and a Google Pixel 3 as a receiver. Transmitter and receiver were placed in a height of $\SI{67}{cm}$ in an indoor environment, and their distance was changed for each measurement. As can be seen, for a given distance, the RSSI differs significantly among the 3 channels. 
In particular, we could observe differences of up to $\SI{15}{dB}$. 
This occurred even for distances below $\SI{2}{m}$, which are the most relevant ones for contact tracing. 

In addition to measured RSSI values, \reffig{fig:RecordedDataDiffDistances} also depicts the path loss model as given by Equation \eqref{eq:pathloss}, 
where $G_t$, $G_r$ and $P_t$ have been fitted to minimize the squared error between model and measurements. 
As already mentioned, the distance is usually estimated based on such a model, which assigns a distance to each RSSI value. 
When now estimating the distance based on the RSSI, an attenuation of $\SI{15}{dB}$ due to the channel-dependent multipath propagation can lead to distance estimation errors in the order of tens of meters. 
Hence, it is of crucial importance to be aware of the channel on which a packet was received. 
If the channel was known, the estimation error due to frequency-selective hardware and free-space path loss could be cancelled using calibration. Furthermore, errors due to multipath propagation could be reduced by computing the average of the same number of RSSI values from different channels, or by advanced methods, e.g.~\cite{ChannelSLAM06}.\\

\par\noindent\textbf{Unavailability of Channel Information: } The BLE radio does not relay the information on which channel a beacon was received to the smartphone operating system. Indeed, the BLE \textit{host control interface}, which is used for data exchange between radio and smartphone, specifies that incoming advertising packets are reported to the host without any channel information~\cite{bleSpec52}. As a result, the frequency-dependent error cannot be mitigated.
A potential method for reducing the error is averaging over multiple packets from different channels. However, without knowing on which channel a packet was received, it cannot be ensured that the same number of packets from each channel are used for averaging.\\

\par\noindent\textbf{Proposed Solution: }In this paper, we for the first time propose a technique to detect on which channel an advertising packet was received on an Android smartphone. We thereby exploit an undocumented behavior of many Bluetooth SoCs found in recent smartphones. In particular, consecutive reception windows are separated from each other by a period of time that is significantly larger than the distance between two consecutive packets. After scanning for incoming packets is activated by an application (app), we observe that on most smartphones we have tested, scanning starts on the same channel. Hence, we can exploit this behavior for obtaining the first channel on which the smartphone scans. We can then classify the channel of later received packets based on associating their reception time with the time windows during which the receiver listens for incoming packets and the corresponding channels. 
\par\noindent\textbf{Contributions: }
Compared to existing works, we in this paper make the following contributions:
\begin{enumerate}
    \item We for the first time propose a technique to identify the channel on which a BLE advertising packet was received.
    \item We evaluate the detection probability of our proposed method in detail. Our results suggest that the channel of reception can be detected reliably in $\SI{100}{\percent}$ of all attempts.
    \item We test our proposed methodology on different smartphone models from different manufactures and show that it is compatible with the vast majority of the phones we have tested.
\end{enumerate}
Besides contact tracing, our proposed method can also be used to improve distance estimation in a host of different applications.\\

\noindent\textbf{Paper Organization: }The rest of this paper is organized as follows. In the next section, we give a brief overview on related work. In \refsection{sec:ble}, we describe how the procedure used for contact tracing in BLE, called \textit{advertising and scanning}, works. Next, in \refsection{sec:channel_detection}, we describe our proposed technique for channel detection. We experimentally evaluate this technique in \refsection{sec:evaluation} and conclude our findings in \refsection{sec:concluding_remarks}.

\section{Related Work}
\label{sec:related_work}
In this Section, we briefly summarize related work on contact tracing and distance estimation using smartphones.\\

\noindent\textbf{Contact Tracing: }
Prior to the global SARS-CoV-2 outbreak in 2020, digital contact tracing did not receive much attention from the scientific community. Early approaches where based on off-the shelf sensor nodes worn by the subjects~\cite{hashemian:10}, or on indoor localization using RFID readers~\cite{hellmich:17}. Research interest massively increased after the beginning of the surge of COVID-19 infections in 2020, and electronic contact tracing is actively being studied today. It appears to be a promising tool for slowing down the spread of SARS-CoV-2, especially because the time window within which contact persons need to be identified must be short for controlling the further spread of the virus~\cite{ferrettie:20}. Since a solution had to become effective  quickly after the outbreak, it was not possible to design and deploy a custom tracing device from scratch. Smartphones were already widely available in the population, and contact tracing could be realized once a suitable application (app) had been developed. However, smartphones were not designed for the purpose of contact tracing~\cite{kindt:20smartphones}. Therefore, their capabilities for contact tracing are limited, which reduces the accuracy and reliability of the tracing procedure. Nevertheless, besides the deployment of multiple national tracing apps all over the world, also the scientific community is working on the design of tracing apps~\cite{gnther2020tracing, wolisz:20, holzapfel:20}. Hence, despite not being the optimal platform, significant effort is being invested in implementing contact tracing on smartphones and improving its performance. Besides privacy, accurate distance estimation, which has already been studied earlier in the context of other applications than contact tracing, remains a central problem. We next describe approaches on distance estimation using smartphones.\\

\noindent\textbf{Distance Estimation using Smartphones: }
Estimating the distance between a pair of smartphones has been studied using different techniques, such as correlating the measurements of the ambient magnetic field by different smartphones~\cite{jeong:19}, acoustic ranging~\cite{thiel:12} or the RSSI of the Bluetooth or IEEE 802.11 (WiFi) radio.
Among these approaches, RSSI-based mathods have turned out to be the most practical one.
Since \ac{BLE} is only available in relatively recent smartphones, early studies have focused on distance estimation using WiFi or the ``legacy'' Bluetooth, called \textit{Bluetooth BR/EDR}. For example, \textit{Comm2Sense}~\cite{carreras:12} configures smartphones as mobile WiFi-hotspots to estimate the proximity of two devices based on the WiFi RSSI. Since setting up mobile WiFi-hotspots on smartphones is inconvenient for the user and the energy consumption of WiFi drains the battery quickly, a large number of other approaches, e.g.,~\cite{liu:14,palaghias:15}, are built on the RSSI of Bluetooth BR/EDR. 
Whereas the RSSI in Bluetooth BR/EDR is measured relatively to a ``golden receive power range''~\cite{bleSpec52}, the BLE protocol specifies that the RSSI is an absolute received power with an accuracy of $\pm\SI{6}{dB}$. In addition, BLE is designed for scanning for incoming beacons continuously in the background. As a result, more recent approaches on distance estimation build upon the BLE protocol. For example, ~\cite{chowdhury:15} uses 3 different approximation models, which are selected based on a coarse classification of the estimated distance. 
A large number of approaches, e.g., \cite{kim:15, ng:19,bisio:16}, have studied the analysis of the sensed RSSI data. Furthermore, the work in \cite{leith:20, leith:20b} has experimentally evaluated the accuracy of distance estimation using smartphones in the context of contact tracing. However, the accuaracy of the RSSI data itself has always been considered as immutable. Hence, to the best of our knowledge, detecting the channel on which a packet was received for improving the accuracy of distance estimation has not been considered previously to this work.

\section{Advertising and Scanning in BLE}
\label{sec:ble}
\begin{figure*}[t!]
	\centering
	\includegraphics[width=\linewidth]{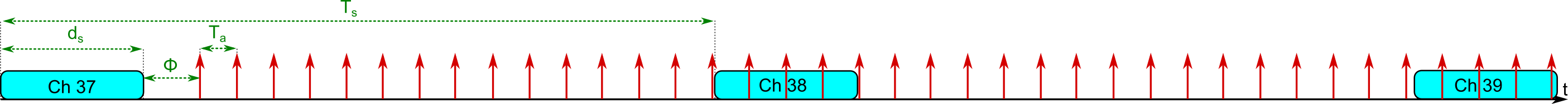}
	\caption{Advertising and scanning in BLE. Arrows depict advertising events, which consist of a sequence of 3 packets on 3 different channels. The rectangles depict scan windows.}
	\label{fig:ble_principle} 
\end{figure*}
A contact tracing app will use the following procedure for detecting devices in its surrounding. It is provided by the BLE protocol~\cite{bleSpec52} and referred to as \textit{advertising and scanning}.

Every device periodically schedules an \textit{advertising event} once per $T_a$ time-units. $T_a$ is called the \textit{advertising interval} and is composed of a static part $T_{a,0}$ plus a random delay $\rho \in [0, \SI{10}{ms}]$. In each such event, three beacons in a row are sent. The first of them is sent on channel 37 (which corresponds to a center frequency of $\SI{2.402}{GHz}$), the second one on channel 38 ($\SI{2.426}{GHz}$) and the third one on channel 39 ($\SI{2.480}{GHz}$)~\cite{bleSpec52}. This is depicted in \reffig{fig:ble_principle}. Here, every arrow stands for an advertising event. Since the 3 individual beacons in each event are sent within short amounts of time, we do not show them in the figure. 

For being able to receive incoming beacons, every device also listens to the channel by using so-called \textit{scan windows}. Every scan window has a duration of $d_s$ time-units, and there is one such window every \textit{scan interval} $T_s$. After every instance of the scan window, the channel for the succeeding window is toggled between channel 37, 38 and 39 in a round-robin fashion. This is also depicted in \reffig{fig:ble_principle}.

A device can detect the presence of another device, once a beacon from the remote device coincides with one of its scan windows~\cite{kindt:18}. Most values for $(T_a, T_s, d_s)$ supported by Android fulfill $T_a < d_s$ (cf. \reffig{fig:ble_principle}) and hence, the reception of at least one beacon is guaranteed in each scan window.
Note that the number of coinciding beacons might vary for every scan window. For example, in \reffig{fig:ble_principle}, the scan window on channel 38 has 3 coinciding beacons, whereas the window on channel 39 has 4.

The Android operating system does not allow an app to select these parameter values directly. Instead, an app can chose between three different settings that determine $T_a$ and three different ones that determine $T_s$ and $d_s$. These settings are listed in \reftable{tab:settings_android}.

It needs to be mentioned that there is no transparent mapping between these settings and the corresponding values. First, the values that correspond to a certain setting (e.g., SCAN\_MODE\_BALANCED) are not officially specified by Google. We have therefore obtained them from the source-code of the latest version of Android\footnote{In older Android versions, the parameter values of the scan modes are different.}. Second, the values that are actually used could differ from those given in \reftable{tab:settings_android} due to scheduling conflicts. In particular, the radio might maintain other Bluetooth connections, and the points in time at which other packets are exchanged might overlap with those needed for advertising and scanning. In addition, the Bluetooth radio is also used for WiFi on many devices, which could lead to additional scheduling conflicts.
However, we found in our experiments that the values from \reftable{tab:settings_android} are actually used during normal operation, i.e., when no scheduling conflicts are present.
 \begin{table}[tbh]
    \centering
 	\begin{tabular}{l|c|c|c}
 		\textbf{Android Setting} &$\mathbf{T_{a}\ [s]}$& $\mathbf{T_s\ [s]}$ & $\mathbf{d_s\ [s]}$ \\
 		\hline
 		SCAN\_MODE\_LOW\_POWER & - & $5.120$&$0.512$\\
 		SCAN\_MODE\_BALANCED & - & $4.096$&$1.024$ \\
 		SCAN\_MODE\_LOW\_LATENCY & - & $4.096$&$4.096$ \\
 		ADVERTISE\_MODE\_LOW\_POWER & $1.000$ & - & - \\
 		ADVERTISE\_MODE\_BALANCED & $0.250$  &- & -\\
 		ADVERTISE\_MODE\_LOW\_LATENCY  & $0.100$ & - & - \\ 		
 	\end{tabular}
 	\vspace*{0.3cm}
 \caption{BLE parameterizations in Android.}
 \label{tab:settings_android}
\end{table}

For the sake of completeness, we here also mention three other configuration options. First, the SCAN\_MODE\_OPPORTUNISTIC setting can be used by an app to obtain scan results when scanning has been triggered by a different app, without triggering the scanning itself. Second, a different set of values for $T_s$ and $d_s$ is available when using \textit{batch scanning}, where multiple received packets are reported to the app jointly after some time instead of immediately after discovery. However, we could not find any documentation of this feature and hence did not study it in detail in this paper. Third, Google and Apple are currently drafting their \textit{Exposure Notification} service~\cite{googleApple:20}. Here, an advertising interval of $\SI{200}{ms}$ to $\SI{270}{ms}$ is specified, while no values for $T_s$ and $d_s$ are given. None of these additional options will supersede the need for channel detection, as we propose in this paper. 

\section{Channel Detection}
\label{sec:channel_detection}
In this section, we describe how the channel on which an incoming beacon is received can be detected on a smartphone. As already mentioned, the BLE radio does not relay this information to the smartphone's operating system, since the Bluetooth standard does not specify an interface for this.

According to the BLE specification~\cite{bleSpec52}, the channel on which the radio scans is toggled  after \textit{every} scan window. Thereby, the same order of channels 37, 38, 39, 37,... is always pursued.

Though the \ac{BLE} specification does not specify on which channel the radio has to scan first after its activation, we could observe on different smartphone models (see \refsection{sec:evaluation} and \reftable{tab:testeddevices}) that when scanning is activated, the device will always begin with channel 37. 

Let the point in time at which scanning was activated be given by $t$. Then, incoming packets will only be received on channel $c\in \{37,38,39 \}$, if their reception time falls within a time-interval $I_{c}(k) = [t_{l,c}(k), t_{r,c}(k)], k = 1,2,3,4,...$, with
\begin{equation}
\label{eq:bleReceptionTimes_theory}
    \begin{array}{ll}
    t_{l,c}(k) &= t + 3 \cdot (k-1) \cdot T_s + (c-37)\cdot T_s\\
    t_{r,c}(k) &= t + 3 \cdot (k-1) \cdot T_s + (c-37)\cdot T_s + d_s\,.
    \end{array}
\end{equation}
Equation~\eqref{eq:bleReceptionTimes_theory} directly follows from \reffig{fig:ble_principle}.
Therefore, we can detect the channel on which a beacon was received by classifying the time of each beacon reception into $I_{37}$, $I_{38}$ or $I_{39}$.

The values for $d_s$ and $T_s$ can be obtained from \reftable{tab:settings_android}. However, recall that especially in case of scheduling conflicts, the phone might potentially deviate from this periodic scheme. Though we could observe that the values from~\reftable{tab:settings_android} appear to be used in \textit{most} cases (i.e., when no scheduling conflicts are present), there might potentially be (slight) changes of these parameter values, or even dropped scan windows or beacons. In our experiments, the scan windows always occurred at the expected points in time given by Equation~\ref{eq:bleReceptionTimes_theory}, even when WiFi was activated. However, some of the transmitted beacons were not received, indicating that the scan windows were interrupted on a short-term basis for carrying out WiFi communication. This does not negatively impact channel detection, since the classification of reception times according to Equation~\refeq{eq:bleReceptionTimes_theory} remains unaffected. The same holds true when beacon transmissions are omitted or their transmission times change due to scheduling conflicts.

Because the clocks of the smartphone and Bluetooth radio are not synchronized, they could drift against each other. This might disturb the channel detection 
based on Equation~\refeq{eq:bleReceptionTimes_theory}, since the classification is carried out within an app that relies on the clock of the smartphone, whereas the scan windows are scheduled using the clock of the radio. To mitigate the effects of this, we slightly modify the interval borders of $I_{c}(k)$ from Equation~\ref{eq:bleReceptionTimes_theory} to $\hat{I}_{c}(k) = [\hat{t}_{l,c}(k), \hat{t}_{r,c}(k)], k = 1,2,3,4,...$, with
\begin{equation}
\label{eq:bleReceptionTimes}
    \begin{array}{ll}
    \hat{t}_{l,c}(k) &= 3 \cdot (k-1) \cdot T_s + (c-37)\cdot T_s + \nicefrac{t_g}{2}\\
    \hat{t}_{r,c}(k) &= t + 3 \cdot (k-1) \cdot T_s + (c-36)\cdot T_s - \nicefrac{t_g}{2}\,.
    \end{array}
\end{equation}
Hence, for realizing a more robust detection, we here classify each received packet by into which instance of $T_s$ it falls, even when being received outside of the (estimated) scan window. 
Further, $t_g$ is a guard time to compensate for the clock drift, for which we propose a concrete value in Section~\ref{sec:evaluation}.
In addition, for limiting this drift, we propose to regularly re-start the scanning after a certain period of time. 
We evaluate after which amount of time such a re-start should occur in Section~\ref{sec:evaluation}.

\refalgo{alg:channelDetection} shows the pseudo-code of our proposed algorithm for detecting the channel on which a packet was received.
In order to limit the power consumption of the smartphone, the algorithm starts the \ac{BLE} scanner using the SCAN\_MODE\_LOW\_POWER\ setting (c.f. Line \ref{algo_Init} in \refalgo{alg:channelDetection}). As soon as \ac{BLE} signals are detected, the main part of the algorithm is executed and the \ac{BLE} scanner is re-started using the SCAN\_MODE\_LOW\_LATENCY setting (c.f. Line \ref{algo_start1}-\ref{algo_start2} in \refalgo{alg:channelDetection}). We remember the timestamp $t$ when the scanning was re-started. Note that this timestamp does not perfectly coincide with the actual re-start of the \ac{BLE} scanning, since there might be delays and/or jitter. The effect of such misalignments are mitigated through the guard time $t_g$ in Equation~\eqref{eq:bleReceptionTimes}. All subsequent packets are handled in the inner while loop starting in Line~\ref{algo_while} of \refalgo{alg:channelDetection}. For each such packet, the reception time is computed and $t$ (i.e., the time when scanning was started) is subtracted. The resulting time difference is used by the \textit{ClassifyChannel()}-function, which infers the channel of reception by evaluating Equation~\eqref{eq:bleReceptionTimes}.
After \textit{Max-Scan-Time} (cf. Line~\ref{algo_max_scan_time} in \refalgo{alg:channelDetection}) has passed, the scanning is re-started to limit the clock drift between smartphone and radio, as already explained. The guard time $t_g$ compensates for any clock drift before this re-start.

In the next section, we evaluate the accuracy of channel detection using this algorithm on different smartphone models and for different re-start intervals of the \ac{BLE} scanning procedure.

\begin{algorithm}
 \BlankLine
 \LinesNumbered
 doInit = true\;
 \While{1}
{
   \If{doInit} 
    {
        Scan-Mode = SCAN\_MODE\_LOW\_POWER\nllabel{algo_Init}\;
        Re-Start-Scan()\;
        doInit = false\;
        Channel-Detection = false\;
    }
    \If{\ac{BLE} signals detected}
    {
        Channel-Detection = true\nllabel{algo_start1}\;
        Scan-Mode = SCAN\_MODE\_LOW\_LATENCY\;
        Re-Start-Scan()\;
        t = GetSystemTime()\tcp*{see Eq.~\eqref{eq:bleReceptionTimes}}\nllabel{algo_start2}
    }
    \While{Channel-Detection\nllabel{algo_while}}
    {
      \If{\ac{BLE} signal received}{
        \tcc{classify \ac{BLE} signal into $\hat{I}_{37}$, $\hat{I}_{38}$ or $\hat{I}_{39}$ using Eq.~\eqref{eq:bleReceptionTimes}}
       ClassifyChannel(t, GetSystemTime())\;
       }
    
      \If{GetSystemTime() - t $>$ Max-Scan-Time }{\nllabel{algo_max_scan_time}
        Re-Start-Scan()\;\nllabel{algo_restart}
        t = GetSystemTime() \tcp*{see Eq.~\eqref{eq:bleReceptionTimes}}
       }

      \If{No signals detected}{ \nllabel{algo_noSignal}
            doInit = true\;
            break\;
      }
    }
}
\caption{Android BLE Channel-Detector}
\label{alg:channelDetection}

\end{algorithm}

\section{Evaluation}
\label{sec:evaluation}
In this section, we evaluate the proper functioning of our proposed technique for channel identification and the accuracy of ~\refalgo{alg:channelDetection}. Towards this, we first describe our experimental setup and then give experimental data.
\subsection{Experimental Setup}
In order to evaluate our approach and~\refalgo{alg:channelDetection} on different smartphone models, we have set up the following test environment.
Four Raspberry Pis\footnote{In our experiments with Samsung Galaxy S5
and Xiaomi MI-A2 smartphones, these Raspberry Pis were replaced by Bluegiga BLE112 radios, which carried out the same tasks as the Raspberry Pis.}, denoted as R($i),\ i = 1\dots4$ continuously transmitted \ac{BLE} advertising packets. Each Raspberry Pi was configured to transmit on a different set of channels, viz., R($1$) on channel 37, R($2$) on channel 38, R($3$) on channel 39 and R($4$) on all three channels, as also shown in \reftable{tab:raspberry}. The advertising interval was set to $\SI{100}{ms}$, which is way below $d_s$ (cf. Table~\ref{tab:settings_android}). Hence, multiple advertising packets will fall into every scan window. Note that $T_a = \SI{100}{ms}$ is used by the ADVERTISE\_MODE\_LOW\_LATENCY setting on Android smartphones. 
Therefore, though having been obtained using Raspberry Pis as senders, our results remain valid when packets are sent using smartphones.
It is worth mentioning that we have tested our proposed methodology also with all other advertising intervals supported by Android (viz., $\SI{100}{ms}$ and $\SI{1000}{ms}$), and found it to be working successfully irrespective of the value of $T_a$ used. 

In BLE, the services a device offers are advertised with the payload of its packets. Such services are identified by unique identifiers, so-called \textit{\acp{UUID}}. We have assigned a different \ac{UUID} to the packets of each Raspberry Pi. 

The packets of $R(4)$, which are sent on all three channels, are received in every scan window. This allowed us to detect the individual scan windows of the smartphone. In addition, we could directly see on which channel a beacon was received, since we have encoded the channel information via the \acp{UUID} of each device. In other words, since e.g., R($1$) only advertises on channel 37 and we can detect that a packet was sent by R($1$) using its \ac{UUID}, we could easily infer on which channel the scanner was scanning.

 \begin{table}[bth]
 \centering
 	\begin{tabular}{c|c|c}
  		\textbf{Raspberry Pi} & Channel & Center Frequency in [GHz]\\
 		\hline
 		R(1) & 37 & $2.402$\\
 		R(2) & 38 & $2.426$\\ 
 		R(3) & 39 & $2.480$\\ 
 		R(4) & 37,38,39 & $2.402,2.426, 2.480$
 	\end{tabular}
 	\vspace*{0.3cm}
 \caption{Experimental setup with 4 Raspberry Pis advertising on different channels.}
 \label{tab:raspberry}
\end{table}

\subsection{Results}
\subsubsection{Behavior of the BLE Radio}
\begin{figure}[tb]
 \begin{subfigure}[b]{\columnwidth}
  \includegraphics[width=0.99\linewidth]{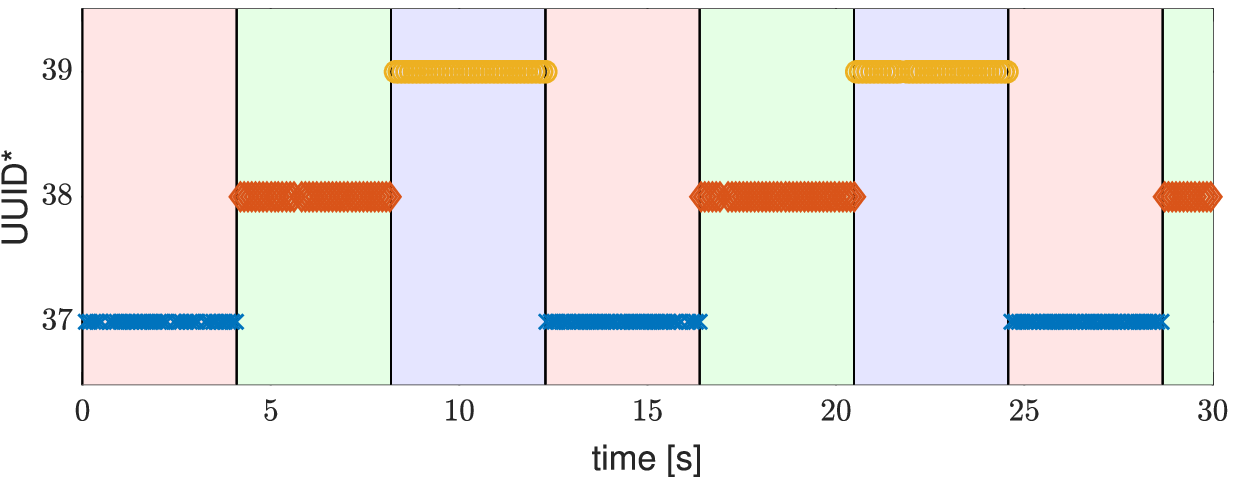}
    \caption{SCAN\_MODE\_LOW\_LATENCY}
    \label{fig:recordingBLEChannel_lowLat}
  \end{subfigure}
   \begin{subfigure}[b]{\columnwidth}
  \includegraphics[width=0.99\linewidth]{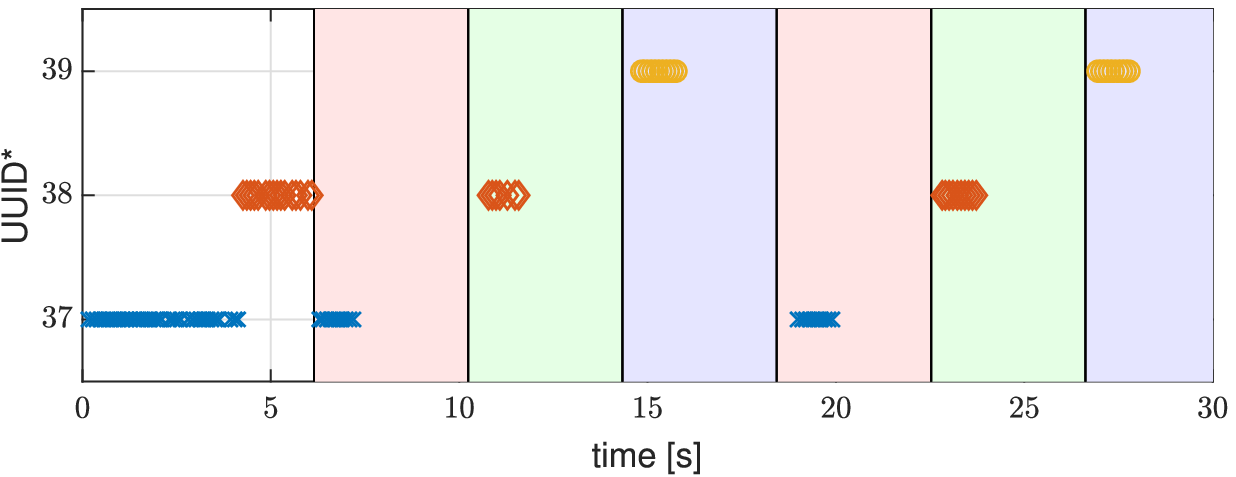}
      \caption{SCAN\_MODE\_BALANCED}
    \label{fig:recordingBLEChannel_bal}
  \end{subfigure}
  
    \caption{Sequence of received packets classified by their channel of reception of the Google Pixel 3 smartphone. The different background colors indicate the estimated channel based on Equation~\refeq{eq:bleReceptionTimes}.}
    \label{fig:recordingBLEChannel}
  \end{figure}

\reffig{fig:recordingBLEChannel_lowLat} shows the reception times of all packets in an experiment using the SCAN\_MODE\_LOW\_LATENCY setting. \reffig{fig:recordingBLEChannel_bal} depicts the results from an experiment in which the SCAN\_MODE\_BALANCED setting was used. Both experiments were carried out using the Google Pixel 3 smartphone.  The individual measurements lie in such a close proximity that they appear as lines rather than as individual points. We have sorted the packets by their \acp{UUID}, which identify their channel.
The different background colors in the figure depict the estimated instances of the scan interval and their channel, as given by Equation~\ref{eq:bleReceptionTimes}. Assuming that scanning started on channel 37, red indicates that in this interval, the scan window was listening on channel 37. Similarly, green corresponds to channel 38 and blue to channel 39. Note that the colors of the depicted received packets do not follow this scheme, since they would be indistinguishable if they had the same color as the background.

 In both experiments, scanning was activated at time $t = \SI{0}{s}$. In the SCAN\_MODE\_LOW\_LATENCY setting, as can be seen in \reffig{fig:recordingBLEChannel_lowLat}, the received packets follow the pattern predicted by Equation~\eqref{eq:bleReceptionTimes}, which is exploited by \refalgo{alg:channelDetection}. In particular, starting from time $t = \SI{0}{s}$, packets are received on channel 37 for $T_s$ time units. Next, a sequence of packets is received on channel $38, 39, 37, 38, 39,\dots$.
 
 In contrast, when using the SCAN\_MODE\_BALANCED setting, we found that the regular scanning pattern begins after a certain offset $t_o$ from the starting time, which is indicated by the white background color in \reffig{fig:recordingBLEChannel_bal}. 
  Before $t_o$ has passed, the smartphone scans using an unpredictable pattern. In the situation exemplified in  \reffig{fig:recordingBLEChannel_bal}, the device first scanned on channel $37$ and then on channel $38$, each for an amount of time that exceeds $d_s$. After $t_o$ has passed, the device re-sets to channel 37 and the scheme as given by Equation~\refeq{eq:bleReceptionTimes_theory} begins. Hence, if the offset $t_o$ would be known, the channel could be classified correctly according to Equation~\eqref{eq:bleReceptionTimes}. However, our measurements with different smartphone models showed that this offset $t_o$ varies every time the scanning is started, and in addition depends on the smartphone model. This justifies that our algorithm is built upon the SCAN\_MODE\_LOW\_LATENCY setting for detecting the channel of an incoming \ac{BLE} signal, despite this setting being the most power-hungry one on Android devices. Since we only switch to this mode for short amounts of time after an initial reception (i.e., until the channel has been identified and a sufficient number of  packets for computing the distance have been received),
the energy overhead of this will be acceptable. Recent results~\cite{kindt:20smartphones} show that - depending on the smartphone model - the battery of the smartphone is drained by between $\SI{5}{\percent}$ and $\SI{20}{\percent}$ earlier compared to when Bluetooth is switched off, if the SCAN\_MODE\_LOW\_LATENCY setting is used during all times. Since we only use this mode for small fractions of the time, the reduction of the battery runtime will be way below this.

 \begin{table*}[tbh]
 \centering
 	\begin{tabular}{c|c|c|c|c|c}
  		\textbf{Device Name}  & \textbf{Model} & \textbf{Compatible} &  \shortstack{\textbf{Classification} \\ \textbf{Accuracy}}   &  \textbf{Android Version} & \textbf{Comments}\\
 		\hline\hline
 		Google Pixel 2 & Pixel 2 &\checkmark & 100 \% & 10 & -\\
 		Google Pixel 3 & Pixel 3 & \checkmark & 100 \%  & 10 & -\\
 		OnePlus 5 & ONEPLUS A5000 & \checkmark & 100 \% & 9 & -\\
 		Samsung Galaxy S5 &  SM-G900F & \checkmark  & 99.5\% & 8.1.0 & Older API: $T_s = \SI{5}{s}$, Lineage OS.\\
 		Xiaomi MI-A2 & M1804D2SG & \checkmark  & 100 \% &  9 & -\\
 		Xiaomi Mi 9T Pro & M1903F11G & \checkmark  & 100 \% &  10 & -\\
 		\hline
 		Huawei P20 lite & ANE-LX1 &\xmark  & - & 9 & Different scan intervals, no channel reset \\
 		Samsung  Galaxy M20 & SM-M205FN &\xmark  & - & 10 & Different scan pattern, no channel reset \\
 		iPhone 6s & MN1E2LL/A & \xmark & - & 13.5.1 (iOS)  & No channel reset
 	\end{tabular}
 	\vspace*{0.3cm}
 \caption{Tested Android devices, a full list of devices can be found in \cite{covidwebpage}\vspace*{-0.5cm}}
 \label{tab:testeddevices}
\end{table*}

\subsubsection{Classification Accuracy}
\begin{figure}[tb]
{
  \centerline{\includegraphics[width=0.99\linewidth]{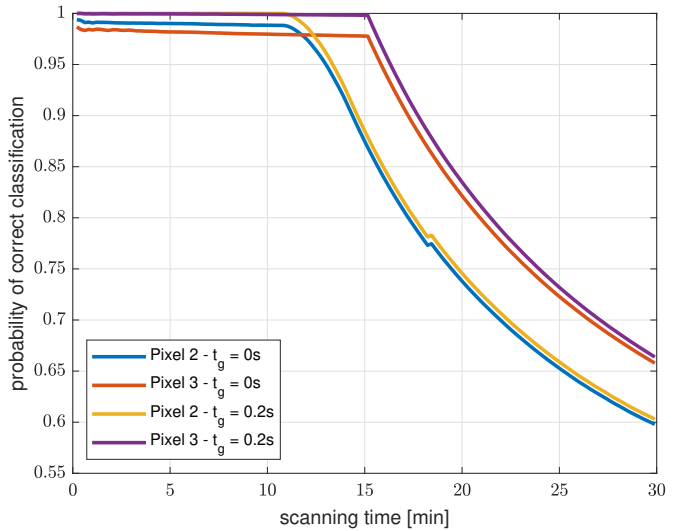}}
    \caption{Probability of classifying the channel correctly (detection accuracy) as a function of the scanning time without re-starting for the Pixel 2 and Pixel 3 smartphones. Two guard times $t_g = \SI{0}{s}$ and $\SI{0.2}{s}$ were applied. The \ac{BLE} scanner was re-started at $t=\SI{0}{s}$.}
    \label{fig:detectionprobability}
 }
 \end{figure}

While~\reffig{fig:recordingBLEChannel} shows that the behavior of the \ac{BLE} radio on multiple smartphone models appears to be suitable for channel detection, in which fraction of all attempts can the channel be classified correctly when using~\refalgo{alg:channelDetection}?
In order to evaluate the success rate of the channel detection algorithm, which we call the \textit{detection accuracy}, \reffig{fig:detectionprobability} shows the fraction of packets for which the channel was detected correctly as a function of the time since the last re-start of the scanning procedure for the Google Pixel 2 and Google Pixel 3 smartphones. Both smartphones continuously recorded the received \ac{BLE} signals of the Raspberry Pis for \SI{24}{h}.
Scanning was re-started every $\SI{30}{min}$. This re-starting was necessary because the Android operating system automatically switches from the SCAN\_MODE\_LOW\_LATENCY to the SCAN\_MODE\_OPPORTUNISTIC setting after $\SI{30}{min}$ of continuous scanning. In the SCAN\_MODE\_OPPORTUNISTIC setting, the device only schedules scan windows when a different app explicitly triggers the scanning (i.e., by using a different mode than SCAN\_MODE\_OPPORTUNISTIC), which would have interrupted our measurements.

We used a guard time of $t_g = \SI{0}{s}$ and $t_g = \SI{0.2}{s}$, respectively. 
As can be seen in the figure, by using a guard time of $t_g = \SI{0.2}{s}$, we obtain a detection accuracy of $\SI{100}{\%}$ for more than $\SI{10}{min}$ of contiguous scanning without re-starting for the Pixel 3 smartphone, and $\SI{15}{min}$ for the Pixel 2 smartphone. For a guard time of $t_g = \SI{0}{s}$, the initial detection accuracy is slightly reduced to around $\SI{97}{\percent}$. When scanning is carried out for more than $\SI{10}{min}$ without reset, the detection accuracy gradually becomes smaller for both values of $t_g$. This is caused by clock drift between the smartphone and radio, as already described. We therefore propose to set $t_g$ to $\SI{0.2}{s}$ and \textit{Max-Scan-Time} in \refalgo{alg:channelDetection} to $\SI{10}{min}$. In general, for a larger value of $t_g$, \refalgo{alg:channelDetection} can be run for a longer time without the need of a reset of the scanning procedure. On the other hand, an increasing number of packets need to be discarded, if $t_g$ is increased.

\subsubsection{Different Smartphone Models}
We have shown that our proposed algorithm works in principle, but will it work for \textit{all} smartphone models from different manufacturers? For answering this question, we have tested different smartphone models for their compatibility with our proposed methodology. In particular, we have tested whether they always start scanning on channel 37 and then switch to the next channel after every instance of $T_s$. We also evaluated the detection accuracy for each of them.
\reftable{tab:testeddevices} summarizes the results of this experiment. Out of 9 smartphones from different manufacturers we have tested, \refalgo{alg:channelDetection} is compatible with 6.

\reftable{tab:testeddevices} shows the classification accuracy for a recording time of \SI{7}{min} and a guard time of $t_g = \SI{0.2}{s}$, similar to \reffig{fig:detectionprobability}. Here, we re-started the scanning after every minute, since we feel that a $\SI{1}{min}$ time-frame is practical for estimating the distance. We obtained a classification accuracy of $100\%$ for 5 of the 6 compatible devices, while the accuracy for the remaining smartphone (viz., the Galaxy S5) is only reduced by $\SI{0.5}{\percent}$. These results can be further improved when tweaking the parameters $t_g = \SI{0.2}{s}$ and the interval after which the scanning is re-started individually for each smartphone model. Also, for most smartphones, a classification accuracy of $\SI{100}{\percent}$ can be maintained for a longer duration than 1 minute (cf. \reffig{fig:detectionprobability}). 

Below, we discuss multiple smartphone models that exhibited an abnormal behavior in detail.\\

\noindent\textbf{Samsung Galaxy S5: } 
\begin{figure}[tb]
\begin{subfigure}[b]{\columnwidth}
\includegraphics[width=0.99\linewidth]{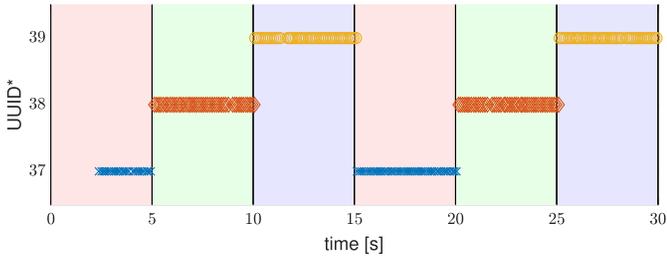}
\caption{Samsung Galaxy S5}
\label{fig:recSamsungHuawei_samsung}
\end{subfigure}
\begin{subfigure}[b]{\columnwidth}
\includegraphics[width=0.99\linewidth]{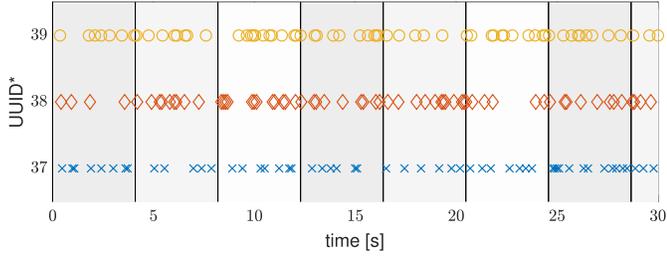}
\caption{Huawei P20 Lite}
\label{fig:recSamsungHuawei_huawei}
\end{subfigure}
\caption{Sequences of received packets classified by their channel of reception over time in the SCAN\_MODE\_LOW\_LATENCY setting for the Samsung Galaxy S5 and the Huawei P20 Lite smartphones. The different background colors in the upper part of the figure indicate the estimated channel of the scan intervals given by Equation~\refeq{eq:bleReceptionTimes}.}
\label{fig:recSamsungHuawei}
\end{figure}
The Samsung Galaxy S5 we have tested was using an older version of Android, in which the scan intervals that were actually used differ from the ones in the most recent Android version. Nevertheless, our proposed algorithm works successfully when adjusting $T_s$. The results of this experiment are shown in \reffig{fig:recSamsungHuawei_samsung}. 
Similarly to \reffig{fig:recordingBLEChannel}, the received packets over time are classified by their channel of reception in \reffig{fig:recSamsungHuawei_samsung}. In our experiment, the SCAN\_MODE\_LOW\_LATENCY setting was used. As can be seen, when adjusting the scan interval, the channel can be tracked reliably over time.\\

\noindent\textbf{Huawei P20 Lite}: 
The Huawei P20 lite and the Samsung Galaxy M20 smartphones were the only two Android smartphones we have tested, on which our proposed methodology did not work. 
In our experiments, we found that the Huawei P20 Lite smartphone used a scan window between $\SI{100}{ms}$ and $\SI{200}{ms}$. 
\reffig{fig:recSamsungHuawei_huawei} depicts the results of this experiment with a Huawei P20 smartphone, in which the  SCAN\_MODE\_LOW\_LATENCY setting was used. As can be seen, the channel is switched much more frequently than expected. However, we could not identify a predictable toggling behaviour of the the Huawei P20 smartphone in our experiments.  
Hence, the algorithm was not able to classify the \ac{BLE} signals into $\hat{I}_{37}$, $\hat{I}_{38}$ or $\hat{I}_{39}$ using Equation~\eqref{eq:bleReceptionTimes}. \\

\noindent\textbf{Samsung Galaxy M20: } 
\begin{figure}[tb]
{
  \centerline{\includegraphics[width=0.99\linewidth]{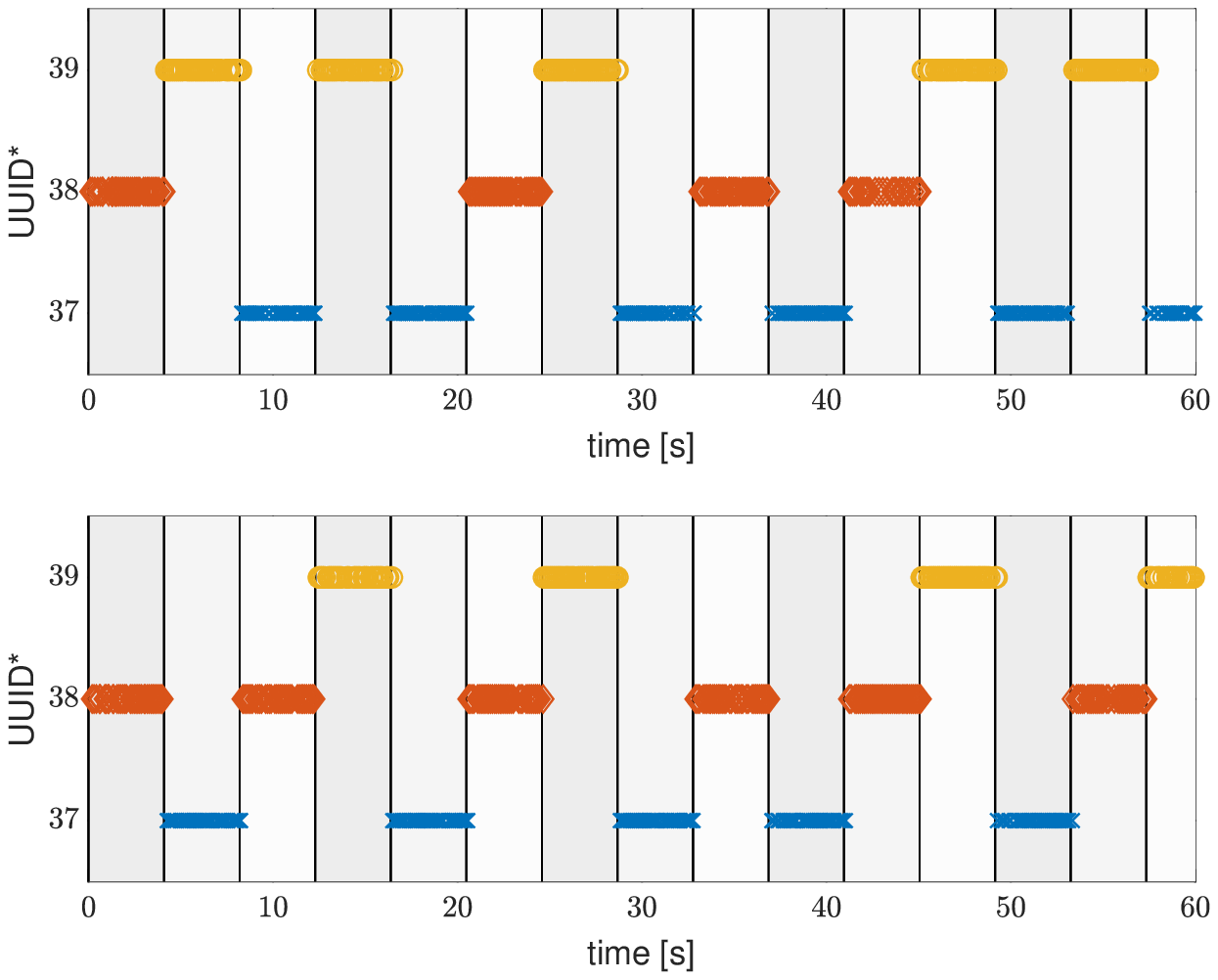}}
    \caption{Two different sequences of received packets classified by their reception channel over time for the Samsung  Galaxy M20 smartphone. Here, the SCAN\_MODE\_LOW\_LATENCY setting was used.
    }
    \label{fig:recordingSamsung}
 }
 \end{figure}
\reffig{fig:recordingSamsung} depicts the received packets classified by their reception channels in two different experiments using the Samsung Galaxy M20 smartphone. 
The Samsung Galaxy M20 smartphone also uses scan windows of length $d_s = \SI{4.096}{s}$ in the SCAN\_MODE\_LOW\_LATENCY setting. However, as can be seen, the channels are toggled in a different order than specified by the BLE specification. For example, in the upper part of the figure, the device started scanning on channel 38, then switched to channel 39, then to channel 37 and then to 39 again. The lower part of the figure depicts another such abnormal scanning sequence. In addition to this abnormality, the starting channel is not reset when the scanning procedure is re-started. Clearly, this scanning pattern does not comply to the \ac{BLE} standard.\\

\noindent\textbf{Apple iPhone 6s: } We also wrote a dedicated Apple iOS app for testing our approach on such devices. Using this app, we could reveal the following behavior on an iPhone 6s. After resetting the scanning, the device always continues to scan on the channel on which it had scanned at the time it was stopped. Clearly, this prevents using our proposed methodology. In addition, the scan windows were scheduled in an unpredictable manner in our experiments. In particular, the scan interval appeared to increase over time. Furthermore, unlike on Android, iOS does not allow an app to choose among different settings that determine $T_s$ and $d_s$. Finally, transmitting in the background, i.e., when the screen is locked, is not possible on iOS.\\

\noindent\textbf{Other Phone Models: } 
In addition to the evaluation data contained in this paper, we plan to test additional smartphone models for their scanning behavior in the future. For this reason, we have launched a website~\cite{covidwebpage} on which we will list the outcomes of additional experiments. The website is available under~\url{https://www.dlr.de/kn/covid}.


\section{Concluding Remarks}
\label{sec:concluding_remarks}
As required by the \ac{BLE} specification, the information on which wireless channel a beacon has been received is not made available to a smartphone. Fortunately, as we have shown in this paper, the behavior of \ac{BLE} radios used on many recent smartphone models allows for reliably detecting the channel indirectly, since the radio always starts scanning channel 37 after a re-start. Hence, we in this paper for the first time proposed a solution to reliably detect the wireless channel on which a packet has been received on a smartphone. We also showed that the channel on which the smartphone scans can be tracked over time by an Android application after re-starting the \ac{BLE} scanning procedure. The proposed technique was experimentally evaluated on multiple different smartphone models. In particular, we showed that a probability of detecting the correct channel of \SI{100}{\%} can be obtained for the Google Pixel 2 and Pixel 3 smartphones, and similar probabilities for other smartphone models.

The information on which channel a packet has been received is highly relevant for distance estimation. e.g., in the context of contact tracing. In particular, the error induced by the frequency-dependent gains $G_r$ and $G_t$, as well as the frequency dependent signal propagation in free space, can be cancelled easily. Also the issue of multi-path propagation, which can occur especially in indoor environments, can be mitigated when the channel on which a packet was received is known. 

Since smartphones have not been designed for contact tracing, tracing apps need to be regarded as a best-effort approach without any guarantees on the performance. For this reason, also a technique that improves the accuracy of distance estimation for \textit{most} (but not all) smartphone models 
will contribute to a higher rate of successfully classified contacts. Therefore, we believe that our proposed technique will be helpful for distance estimation in contact tracing apps against COVID-19. Besides this, a host of other distance estimation-based applications can benefit from our proposed technique.

\section{Acknowledgement}
The authors would like to thank Prof. Dr. Christoph G\"unther, Dr. Armin Dammann and Prof. Dr. Samarjit Chakraborty for many interesting discussions on this paper.

\bibliographystyle{IEEEtran}
\bibliography{./literature}

\end{document}